\documentclass[12pt]{iopart}
\usepackage{epsfig} 

\begin{document}

\title{How to focus a Cherenkov telescope}
\author{W. Hofmann}
\address{Max-Planck-Institut f\"ur Kernphysik, P.O. Box 103980, 
D 69029 Heidelberg}
\ead{werner.hofmann@mpi-hd.mpg.de}

\begin{abstract}
Cherenkov telescopes image the Cherenkov emission from air showers.
A priori, it is not obvious if the `best' images are achieved by
measuring Cherenkov photon angles, i.e. focusing the telescope
at infinity, or by considering the air shower as an object to be
imaged, in which case one might focus the telescope on the central region
of the shower. The issue is addressed using shower simulations.
\end{abstract}

\submitto{\JPG}

\maketitle

\section{Introduction}

Imaging Atmospheric Cherenkov Telescopes (IACTs) have emerged as the most successful
instrument for gamma-ray astrophysics in the TeV energy range
(e.g., Weekes 1996). IACTs image the Cherenkov
light emitted by an air shower onto a highly sensitive photon detector
(the `camera'), typically
a matrix of photomultiplier tubes (PMTs). Viewed with a telescope located at a 
distance of O(100 m) from the shower axis, the air shower generates an elongated
image. The orientation of the image reflects the orientation of the shower axis,
the intensity of the image the shower energy, and the width of the image relates
to the shower type and can be used to distinguish gamma-induced electromagnetic
showers and hadronic background showers. Combining the multiple views provided by
stereoscopic systems of Cherenkov telescopes, the shower axis can be unambiguously
located in space; typical event-by-event precision is $0.1^\circ$ for the direction
of the shower axis, and $\approx 10$~m for the location of the shower impact point
(Aharonian et al. 1997, Konopelko et al. 1999).

The optical element of IACTs is a large reflector, usually composed of a number
of smaller mirror segments. With a shower light yield of about 100 
photons~m$^{-2}$TeV$^{-1}$
and a typical photodetector efficiency around 10\%, mirror areas of about 10 m$^2$ and
100 m$^2$, respectively, are required to provide reasonable images
(O(100) detected
photons) of 1 TeV and 100 GeV gamma-ray showers. With such large mirrors,
the shower image has a limited depth of field, and one needs to decide how to
focus the telescope:
\begin{itemize}
\item One option is to image photon directions, i.e. focus the telescope at 
infinity by locating the camera at a distance $f$ from the mirror, where $f$
is the focal length. In the small-angle approximation, 
the (`plate')coordinate $\vec{q}$ of a photon detected in the 
camera plane measures the slopes $\vec{\theta} = (\theta_x,\theta_y)$ 
of the photon direction relative to the telescope axis (which defines the 
$z$ direction)
\footnote{In the convention used here, the mirror is treated like a
lens, with the `object' at $-z$ and the image at $+z$.}
$$
\vec{q} = f \vec{\theta}
$$
In the following, plate coordinates $\vec{p} = (x,y)$ will be expressed
in units of degrees, dividing out the factor $f$:
$$
\vec{p} = {180^\circ \over \pi}~\vec{\theta}
= {180^\circ \over \pi}~{\vec{q} \over f}~~~.
$$
Since the direction of Cherenkov photons - rather than their location - 
relates to the direction of the primary, this may seem a natural choice.
\item Alternatively, one might consider the IACT taking a photograph of
the air shower, in which case one will focus the telescope at the typical
distance $S$ between the air shower and the telescope. 
This is achieved by locating the camera at a distance 
$f/(1-f/S)$ from the mirror. After appropriate rescaling of plate
coordinates to achieve the same magnification, one finds
$$
\vec{p} = {180^\circ \over \pi}~(\vec{\theta} - \vec{r}/S)
$$
where $\vec{r}$ is the point where the photon hits the mirror, relative
to the center of the mirror.
\end{itemize}
A priori, it is not clear if this choice matters, and if yes, which
choice is best. This note aims to settle the issue, based both on
simple analytical arguments and on simulations.

\section{Width of shower images}

I will argue that the most important criterion is the observed width
of the shower image, as compared to an ideal optical system and detector,
i.e. a very small mirror with very large depth of field and a camera
with very fine pixels. The width of the image is important both for
the separation between narrow gamma-induced showers and wide nuclear
showers, and for the angular resolution of the instrument. 
Any broadening
of the image will worsen the cosmic-ray
rejection capability of the instrument. Wider images will also make it more
difficult to accurately determine the image axis, and hence the direction
of the primary. In optimizing the optics, one can therefore simply aim at 
generating the narrowest images for gamma-ray primaries, provided
that there are no other side effects. Since the
reconstructed image represents a convolution of the ideal image with
the instrumental response, instrumental effects can only reduce
the separation capabilities between gamma-ray showers and nucleon-induced
shower, and worsen the angular resolution 
\footnote{Strictly speaking, this holds under the 
assumption that the data analysis makes optimal
use of the recorded information.}.

\begin{figure}
\begin{center}
\mbox{
\epsfxsize7.0cm
\epsffile{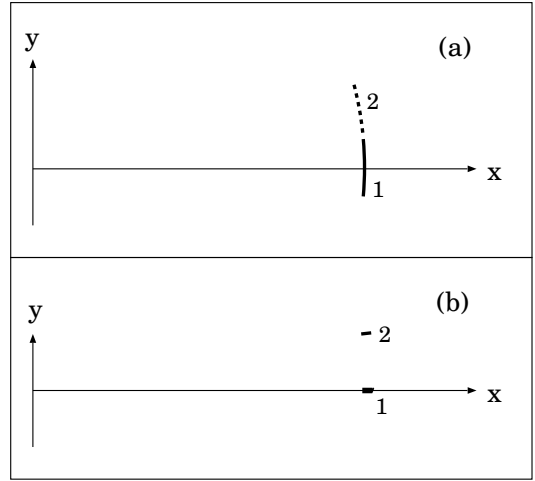}}
\label{focfig0}
\caption{Image generated by a short track segment of a particle
at roughly 10 km height, viewed at a distance of about 120 m
from the shower axis with a telescope with 20 m mirror diameter,
in case the particle moves along the shower axis (1) or is
displaced sideways by 20 m (2). The $x$ axis is defined as
pointing from the shower axis to the telescope location.
(a) For a telescope focused at infinity, and (b) for a telescope
focused at 10 km height.}
\end{center}
\end{figure}
A number of effects will potentially contribute to the reconstructed
width of the shower image in the camera:
\begin{itemize}
\item The transverse radius of the shower $R_{shower}$ at the typical
distance $D$ from the telescope translates into a width 
$\Delta \theta \approx 2 R_{shower} / D $. This dependence is used
to distinguish gamma-induced showers with a characteristic value
$R_{shower} \approx 20$~m and nucleon-induced showers with 
$R_{shower} \approx 70$~m (Hillas 1996). 
With $D \approx 10$~km one finds a width
of $0.2^\circ$ for gamma showers
\footnote{For these rough estimates, we 
use $2 R_{shower}$ as characteristic maximal
distance between the emission points of two photons, and ignore the
exact numerical factors arising from an integration over a circular
emission area, or the actual radial distribution of shower particles.
Equivalent simplifications are used in the following estimates.};
the core region with the high density of Cherenkov photons is quite a bit
narrower. 
\item The finite size $R_{mirror}$ of the mirror implies that the slopes
of photons emitted from an assumed point source at distance $D$
and collected on the mirror vary by 
$\Delta \theta \approx 2 R_{mirror} / D$, when the telescope is
focused at infinity. A 10~m mirror will generate a segment of
$\approx 0.1^\circ$ transverse width
(Fig.~\ref{focfig0}(a)). Focused at $D$, the mirror will generate a point image,
modulo the effects of optical aberrations discussed below
(Fig.~\ref{focfig0}(b)).
\item The size $\delta$ of the camera image elements introduces an
uncertainty $\Delta \theta \approx \delta / f$ in the direction of
detected photons. Minimal pixel sizes used in today's cameras
are $0.1^\circ$; next-generation photon detectors may allow a 
significantly finer segmentation of the cameras.
\item Optical imperfections of the mirror, both due to imperfections
in the manufacturing of the mirrors, and due to unavoidable
optical aberrations in particular for off-axis rays will smear
the image. The aberrations can be reduced by 
increasing the $f/d$ ratio of the mirror; for Davies-Cotton
mirror optics, the dominant aberration is of the scale
$\Delta \theta \approx 0.1^\circ (f/d)^{-2} \theta$ (fwhm), where
$\theta$ is the angle of rays relative to the optical axis,
in units of degrees.
\end{itemize}
One notes that for a large mirror ($R_{mirror} = 10$~m) and a modern
camera (with $0.1^\circ$ pixels) 
the effects of shower size, mirror size, camera pixel size,
and optical aberrations for $f/d \approx 1$
are of similar order of magnitude. Optimal focusing of the mirror
then becomes a relevant issue. Image width will be minimized
by focusing at the height of those shower particles most
relevant for the reconstruction of shower characteristics.
Focusing at the typical shower height rather than at infinity
will also slightly modify the longitudinal profile of the shower,
which is however much less sensitive to imaging properties
since its scale is dominated by the longitudinal evolution of
the shower.

\section{Shower simulations}

To illustrate the effects on the focus distance in more detail,
the distribution of Cherenkov photons in the image plane was
studied in simulations using the CORSIKA air shower code
(Heck et al., 1998), with
modifications by Bernl\"ohr (2000) to include Cherenkov emission.
The simulations were carried out for 100 GeV photon showers
at vertical incidence, with the observation plane located at
2.2 km asl. (the height of the HEGRA installation at La Palma).
A dish diameter of 20 m was assumed. In the simulation code,
a reduced step size for the simulation of multiple scattering
was used, to ensure proper scattering angles between the emission
of successive photons hitting the mirror; with typical default values,
multiple photons emitted from one straight track segment may be
detected and give the (false) impression of nice Cherenkov rings.
The simulations neglect optical aberrations, i.e., assume a 
large $(f/d)$. In the initial simulations, the geomagnetic field
was turned off; geomagnetic fields cause additional distortions
and smearing of the images, as discussed by Chadwick et al. (1999),
and addressed below.

Fig.~2 shows contour lines of the average photon density in images,
viewing the shower at a distance of $R = 120$~m from the axis. The contours
are given for different focus distances, varying between infinity
and 5~km. In the Figure, the $x$ axis corresponds to the major axis
of the images, i.e., to the direction towards the shower axis.
The $x$ coordinate relates to the height $h$ of the photon emission;
$x \approx (180^\circ/\pi) (R/h)$.
The $y$ axis denotes the transverse coordinate, in which the width
of a shower is determined. The influence of the focus distance is
quite dramatic; a focus at 15~km to 20~km significantly narrows
the small-$x$, large-$h$ part of the image, whereas a shorter
focus distance concentrates the tail end of the image.
The pattern is qualitatively the same for different values of
the distance between shower axis and telescope.
\begin{figure}
\begin{center}
\mbox{
\epsfxsize13.0cm
\epsffile{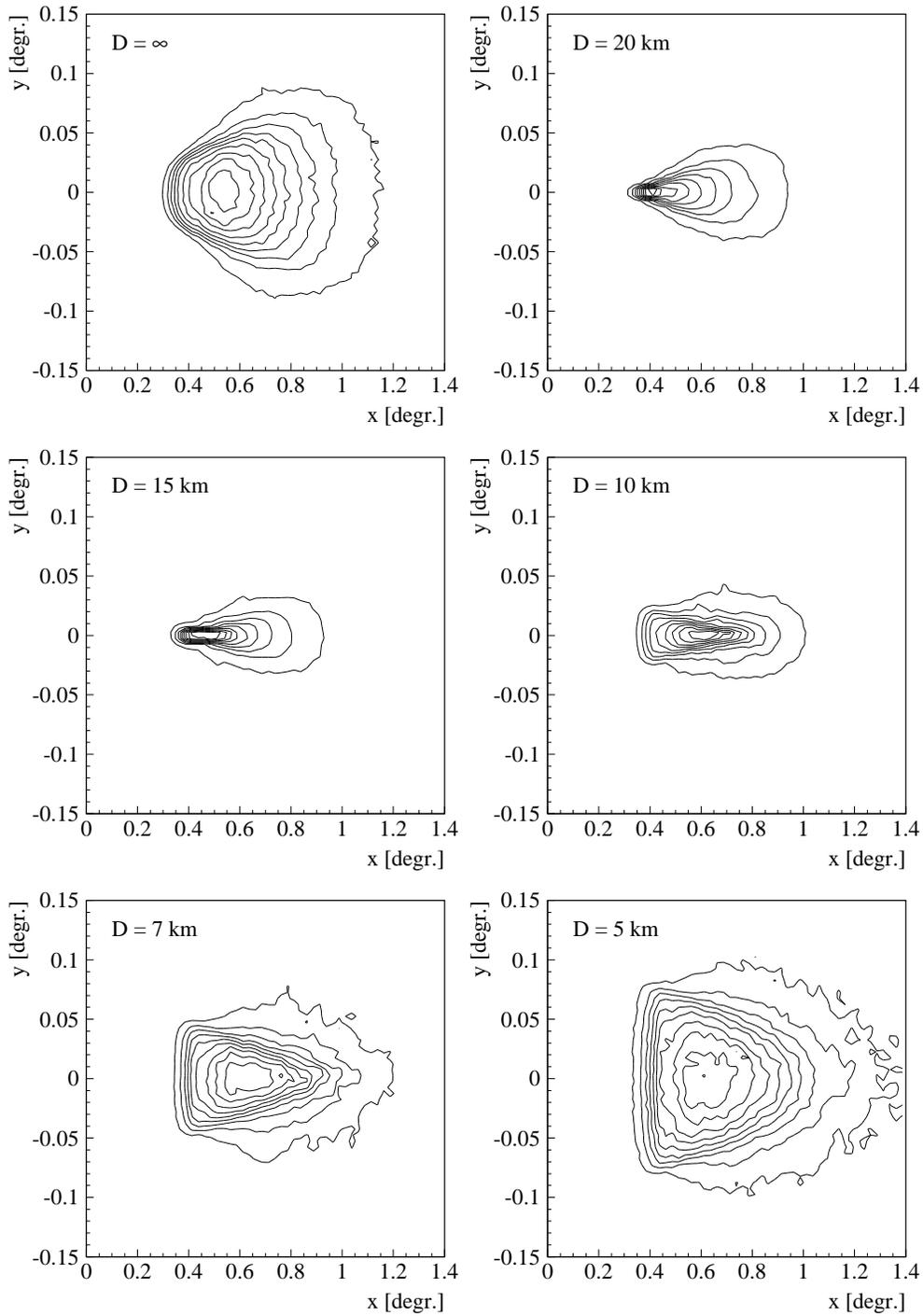}}
\label{focfig1}
\caption{Equidistant contour lines of the photon density
in Cherenkov images of 100 GeV vertical gamma-ray showers, viewed 
at a distance of 120 m from the shower axis with a 20 m
diameter reflector, focused at different distances (infinity,
20 km, 15 km, 10 km, 7 km, 5 km). The observation plane
is located 2.2 km asl. Optical aberrations are 
not included in the simulation. Shower simulations do not
include the effects of geomagnetic fields. 
The $x$ axis is defined as the direction towards to telescope,
i.e., the major axis of images. Note the difference
in scales for the $x$ and $y$ axes. The spacing of contour levels
scales with the peak intensity, and varies between figures.}
\end{center}
\end{figure}

The contour lines in Fig.~2 tend to emphasize the sharp cusp in
the center of the image, and the long tails of the distributions
are not well visible; to provide a more quantitative measure,
the width of images at a given $x$ value was defined as the full
width of the region in $y$, which -- centered at $y = 0$ --
contains 50\% of all photons at that $x$ value. 
The result, see Fig.~3, confirms
the conclusions drawn from Fig.~2. Proper focusing can reduce
the width of the image by almost a factor 10 in certain
$x$-regions, and by about a factor 2 when averaged over $x$.
The simulations show that the
longitudinal profile is virtually unaffected by the choice
of the focal distance.
\begin{figure}
\begin{center}
\mbox{
\epsfxsize10.0cm
\epsffile{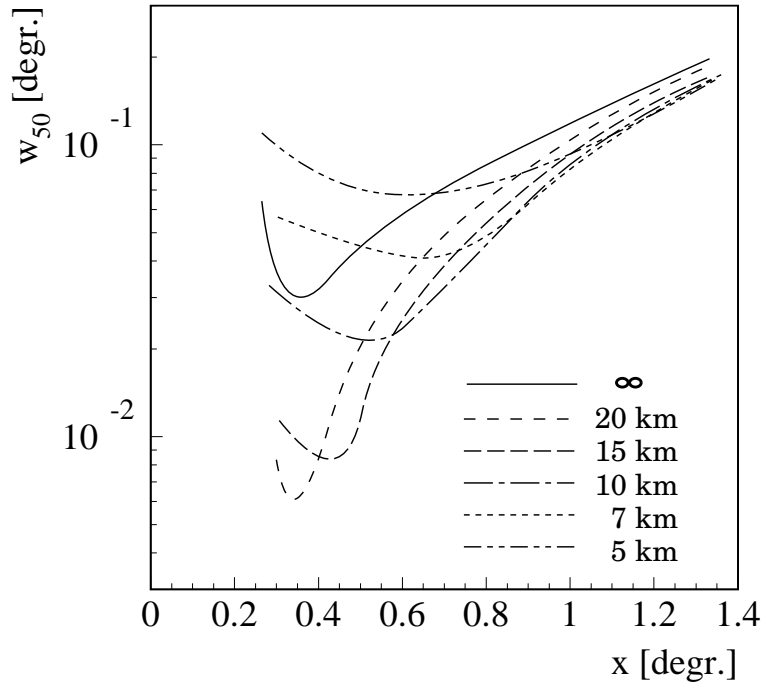}}
\label{focfig2}
\caption{Transverse width of shower images as a function 
of the $x$ coordinate along the major axis of images,
for a reflector focused at distances of infinity,
20 km, 15 km, 10 km, 7 km, and 5 km. The width is defined
as the full width of the range in $y$ which contains 50\%
of all photons, for a given slice of the image.}
\end{center}
\end{figure}

The optimal focus will depend on the other characteristics
of the instrument and on the analysis procedures. For an
instrument with small optical aberrations, i.e. sufficiently
large $(f/d)$ and with very fine pixels, and for high-intensity
images one could use the low-$x$ core region of the images to
obtain a very precise shower direction -- a resolution in
the $0.01^\circ$ range seems feasible. For such an application,
one will use a focus at about 15~km. In telescopes with larger
aberrations and larger pixels, one should instead provide
the optimum focus for the bulk of the image, with a focal
distance around 10~km. Taking into account that most
observations will be performed at non-zero zenith angles,
one might prefer a slightly larger focal distance than
the two values given for vertical showers.

Depending on the location and pointing of the telescopes,
the deflection of shower electrons by the geomagnetic
field may be significant, and result in a distortion and
smearing of the images. The influence of the geomagnetic
field was studied using the La Palma field values; characteristic
image contours are shown in Fig.~4. While the 
images widen somewhat, the conclusions concerning the focusing
of the telescope remain unchanged. In particular, it is still
possible to obtain a very narrow image of the high-altitude
part of the air shower.
\begin{figure}
\begin{center}
\mbox{
\epsfxsize13.0cm
\epsffile{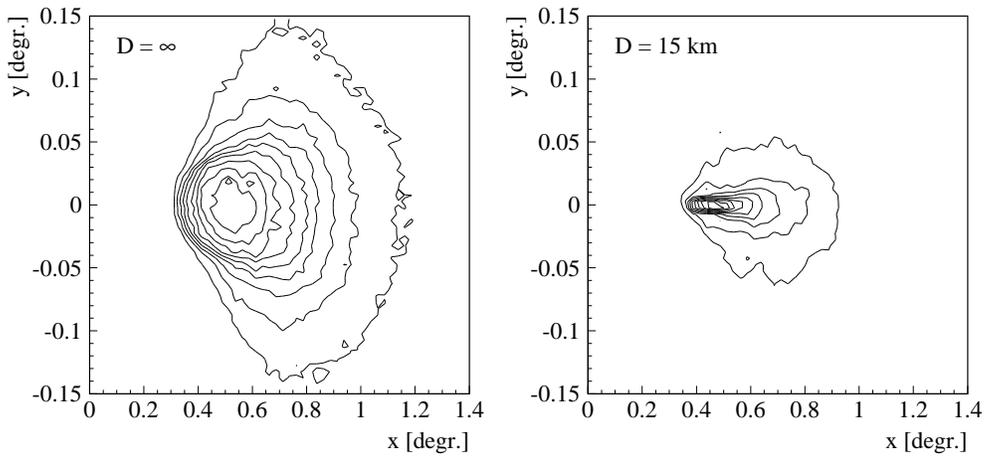}}
\label{focfig3}
\caption{Equidistant contour lines of the photon density
in Cherenkov images of 100 GeV vertical gamma-ray showers, viewed 
at a distance of 120 m from the shower axis with a 20 m
diameter reflector, focused at different distances (infinity,
15 km). Optical aberrations are 
not included in the simulation. Shower simulations include
the geomagnetic field at La Palma.}
\end{center}
\end{figure}

\section{Concluding remarks}  

In particular for next-generation Cherenkov telescopes with
very large mirrors and fine segmentation of the photon detector,
the focusing of the telescope becomes a relevant issue;
this is obviously the case once the diameter of the mirror is
comparable to the transverse size of an air shower.
Optimal imaging of the Cherenkov light from air showers is
achieved by focusing the telescope on the relevant portion
of an air shower. For a classical Hillas-type analysis,
which averages over the longitudinal profile of a shower
when determining the width, one should focus on the region
of the shower maximum. Advanced image analyses resolving fine
details of the image and making use of the sharp core region
of the image are better served by focusing on the head of
the air shower.

\section*{References}
\begin{harvard}

\item[] Aharonian, F.A., Hofmann, W., Konopelko, A.K., V\"olk, H.J.,
1997, {\it Astropart. Phys.} {\bf 6}, 343 and {\it Astropart. Phys.} {\bf 6}, 396
\item[] Bernl\"ohr, K., 2000, {\it Astropart. Phys.} {\bf 12}, 255 
\item[] Chadwick, P.M., et al., 1999, {\it J. Phys.} G {\bf 25}, 1223
\item[] Heck, D., et al., 1998, CORSIKA: A Monte Carlo code to simulate
extensive air shower, FZKA 6019 (Forschungszentrum Karlsruhe)
\item[] Hillas, A.M., 1996, {\it Sp. Sc. Rev.} {\bf 75}, 17
\item[] Konopelko, A., et al., 1999, {\it Astopart. Phys.} {\bf 10}, 275
\item[] Weekes, T.C., 1996, {\it Sp. Sc. Rev.} {\bf 75}, 1

\end{harvard}

\end{document}